\documentclass[12pt]{article}
\usepackage{amsmath}
\usepackage{amssymb,palatino,theorem}
\usepackage{psfig}
\usepackage{verbatim}

\usepackage{graphicx}
\newcommand{\ls}[1]
    {\dimen0=\fontdimen6\the\font \lineskip=#1\dimen0
\advance\lineskip.5\fontdimen5\the\font \advance\lineskip-\dimen0 \lineskiplimit=.9\lineskip \baselineskip=\lineskip \advance\baselineskip\dimen0
\normallineskip\lineskip \normallineskiplimit\lineskiplimit \normalbaselineskip\baselineskip \ignorespaces }

\makeatletter
\def\@begintheorem#1#2{\it \trivlist \item[\hskip \labelsep{\bf #1\
#2.}]} \makeatother

\newcommand{\be}{\begin{equation}}
\newcommand{\ee}{\end{equation}}
\newcommand{\bea}{\begin{eqnarray}}
\newcommand{\eea}{\end{eqnarray}}

\newcommand{\beq}[1]{\begin{equation}\label{#1}}
\newcommand{\eeq}{\end{equation}}

\newcommand{\beqn}[1]{\begin{eqnarray}\label{#1}}
\newcommand{\eeqn}{\end{eqnarray}}

\newcommand{\beaa}{\begin{eqnarray*}}
\newcommand{\eeaa}{\end{eqnarray*}}

\def\SNR    {\mbox{\scriptsize\sf SNR}}

\def \det       {{\rm det}}

\def \nT        {n_\mathrm{\scriptscriptstyle T}}
\def \nR        {n_\mathrm{\scriptscriptstyle R}}

\def \Tr        {\mathrm{Tr}}

\def\Hm{{\bf H}}

\def\AbvGT #1#2{\lower2pt\vbox{\baselineskip0pt \lineskip-.5pt%
         \halign{$#1 ##$\cr #2\crcr >\cr}}}

\def\complex{\mathop{\raise .45ex\hbox{${\bf\scriptstyle{|}}$}
      \kern -0.40em {\rm \textstyle{C}}}\nolimits}
\def\hilbert{\mathop{\raise .21ex\hbox{$\bigcirc$}}\kern -1.005em
{\rm\textstyle{H}}} %Hilbert space
\def\PARstart#1#2{\begingroup\def\par{\endgraf\endgroup\lineskiplimit=0pt}
    \setbox2=\hbox{\uppercase{#2} }\newdimen\tmpht \tmpht \ht2
    \advance\tmpht by \baselineskip\font\hhuge=cmr10 at \tmpht
    \setbox1=\hbox{{\hhuge #1}}
    \count7=\tmpht \count8=\ht1\divide\count8 by 1000 \divide\count7 by\count8
    \tmpht=.001\tmpht\multiply\tmpht by \count7\font\hhuge=cmr10 at \tmpht
    \setbox1=\hbox{{\hhuge #1}} \noindent \hangindent1.05\wd1
    \hangafter=-2 {\hskip-\hangindent \lower1\ht1\hbox{\raise1.0\ht2\copy1}%
    \kern-0\wd1}\copy2\lineskiplimit=-1000pt}

\def\squarebox#1{\hbox to #1{\hfill\vbox to #1{\vfill}}}

\title{\bf Transmit Diversity v. Spatial Multiplexing in Modern MIMO Systems}

\author{Angel Lozano\thanks{Angel Lozano is with Universitat Pompeu Fabra, Barcelona 08005, Spain.
His work is partially supported by the project CONSOLIDER-INGENIO 2010 CSD2008-00010 "COMONSENS".}, and
Nihar Jindal\thanks{Nihar Jindal is with the University of Minnesota,
Minneapolis, MN55455, USA. His work was partially conducted during a visit to UPF under the sponsorship of Project TEC2006-01428.}}

\begin{document}

\maketitle

\begin{abstract}
A contemporary perspective on the tradeoff between transmit antenna diversity and spatial multiplexing is provided.
It is argued that, in the context of most modern wireless systems and for the operating points of interest, transmission
techniques that utilize all available spatial degrees of freedom for multiplexing outperform techniques that
explicitly sacrifice spatial multiplexing for diversity.
In the context of such systems, therefore, there essentially is no decision to be made between transmit antenna diversity and spatial multiplexing in MIMO communication.
Reaching this conclusion, however, requires that the channel and some key system features be adequately modeled and that suitable
performance metrics be adopted; failure to do so
may bring about starkly different conclusions.
As a specific example, this contrast is illustrated using the 3GPP Long-Term Evolution system design.
\end{abstract}

%\bigskip
%{\bf Keywords:} Diversity; Spatial Multiplexing; OFDM; MIMO; DMT; Multiantenna Communication

%\eject

%%%%%%%%%%%%%%%%%%%%%%%%%%%%%%%%%%%%%%%%%%%%%%%%%%%%%%%%%%%%%%%%%%%%%%%%%%%%%%%%%
\section{Introduction}
\label{intro}

Multipath fading is one of the most fundamental features of wireless channels. Because multiple received replicas of the transmitted signal sometimes
combine destructively, there is a significant probability of severe fades. Without any means of mitigating such fading, ensuring reasonable reliability
requires hefty power margins.

Fortunately, fades, or nulls, are very localized in space and
frequency: a change in the transmitter or
receiver location (on the order of a carrier wavelength) or in the
frequency (on the order of the inverse of the propagation delay spread)
leads to a roughly independent realization of the fading process. Motivated by this \emph{selectivity},
the concept of \emph{diversity} is borne: rather than making the success of
a transmission entirely dependent on a single fading realization,
hedge the transmission's success across multiple realizations in
order to decrease the probability of failure.
Hedging or diversifying are almost universal actions
in the presence of uncertainty, instrumental not only in communications but also in other fields as disparate
as economics or biology.

In communications specifically, the term 'diversity' has, over time, acquired different meanings, to the point of becoming overloaded.
It is used to signify:
\begin{itemize}
\item Variations of the underlying channel in time, frequency, space, etc. %We thus speak of "time diversity", "frequency diversity", "spatial diversity", etc.
\item Performance metrics related to the error probability. Adding nuance to the term, more than one such metric can be defined (cf. Section \ref{mainsection}).
\item Transmission and/or reception techniques designed to improve the above metrics.
\end{itemize}
In this paper, we carefully discriminate these meanings.
We use 'selectivity' to refer to channel features, which are determined by the environment (e.g., propagation and user mobility) and by basic system
parameters (e.g., bandwidth and antenna spacing).
In turn, the term 'diversity' is reserved for performance metrics and for specific transmit/receive techniques, both of which have to do with the signal.
Note that channel selectivity is a necessary condition for diversity strategies to yield an improvement in some diversity metric.

\subsection{Diversity over Time}

Archaic electrical communication systems from a century ago already featured primitive forms of diversity,
where operators manually selected the receiver with the best quality.
Automatic selection of the strongest among various receivers was discussed as early as 1930 \cite{bohn1930}.
This naturally led to the suggestion of receive antenna combining, initially for
microwave links \cite{beverage1931}--\nocite{altman1956,bello1963}\cite{makino1967}.
MRC, by far the most ubiquitous combining scheme, was first proposed in 1954 \cite{kahn1954}.
In addition to receive antenna combining, other approaches such as the aforementioned one of repeating the signal on two or more frequency channels
were also considered for microwave links \cite{Barnett1970}. (Systems were still analog and thus coding and interleaving was not an option.)
Given the cost of spectrum, though, approaches that consume additional bandwidth were naturally unattractive and thus
the use of antennas quickly emerged as the preferred diversity approach.
Recognizing this point, receive antenna combining was debated extensively
in the 1950's \cite{brennan1959}--\nocite{schwartz1965,vigants1968}\cite{jakes} and has since been almost universally
adopted for use at base station sites. The industry, however, remained largely ambivalent about multiple antennas at
mobile devices. Although featured in early AMPS trials in the 1970's, and
despite repeated favorable studies (e.g., \cite{cox83}),
until recently its adoption had been resisted.\footnote{The sole exception was the Japanese PDC system \cite{pdc},
which supported dual-antenna terminals since the early 1990's.}

Multiple base station antennas immediately allow for uplink receive diversity. It is less clear, on the other hand, how to achieve diversity in the
downlink using only multiple transmit antennas. In Rayleigh fading, transmitting each symbol from every antenna simultaneously is equivalent to
using a single transmit antenna \cite[Section 7.3.2]{Goldsmith_Wireless}. Suboptimal schemes were formulated that
convert the spatial selectivity across the transmit antennas into effective time or frequency selectivity.
In these schemes, multiple copies of each symbol are transmitted from the various antennas, each subject to either a phase shift \cite{PSTD} or a time delay \cite{wittneben}.
From the standpoint of the receiver, then, the effective channel that the signal has passed through displays enhanced time or frequency selectivity and
thus a diversity advantage can be reaped with appropriate coding and interleaving (cf. Section I-C).

More refined transmit diversity techniques did not develop until the 1990's.
Pioneered in \cite{alamouti},
these techniques blossomed into STBC (space-time block codes) \cite{tarokh} and, subsequently, onto space-time codes at large.
Albeit first proposed for single-antenna receivers,
STBC's can also be used in MIMO (multiple-input multiple-output) communication, i.e., when both transmitter and receiver have a multiplicity of antennas.
This yields additional diversity, and thus reliability, but no increases in the number of information symbols per MIMO symbol.

Concurrently with space-time coding, the principles of spatial multiplexing were also formulated in the 1990's \cite{jerry}--\nocite{jerry2,telatar}\cite{cioffi}.
The tenet in spatial multiplexing is to transmit different
symbols from each antenna, and have the receiver discriminate these
symbols by taking advantage of the fact that each transmit antenna has
a different spatial signature at the receiver (because of spatial selectivity).
This does allow for an increased number of information symbols per MIMO symbol, but it does not enhance reliability.

Altogether, the powerful thrust promised by MIMO is finally bringing
multiantenna devices to the marketplace. Indeed, MIMO is an integral feature of emerging wireless systems such as
3GPP LTE (Long-Term Evolution) \cite{LTEbook}, 3GPP2 Ultra Mobile Broadband, and IEEE 802.16 WiMAX \cite{AndrewsBook}.

\subsection{Overview of Work}

With the advent of MIMO, it may seem that a choice needs to be made
between transmit diversity techniques, which increase reliability (decrease
probability of error), and spatial multiplexing techniques,
which utilize antennas to transmit additional information but
do not increase reliability.
Applications requiring extremely high reliability seem well suited for transmit diversity techniques whereas
applications that can smoothly handle loss appear better suited for spatial multiplexing.
It may further appear that the SNR (signal-to-noise ratio)
and the degree of channel selectivity should also affect this decision.

Our findings, however, differ strikingly from the above intuitions.
The main conclusion is that techniques utilizing all
available spatial degrees of freedom for multiplexing outperform, at
operating points of interest for modern wireless systems, techniques that
explicitly sacrifice spatial multiplexing for transmit diversity. Thus,
from a performance perspective there essentially is no decision that
need be made between transmit diversity and multiplexing in contemporary MIMO systems.
This conclusion holds even when suboptimal spatial multiplexing techniques are used.

There are a number of different arguments that lead to this conclusion, and which will be
elaborated upon:
\begin{itemize}
\item Modern systems use link adaptation to maintain a target error probability and
there is essentially no benefit in operating below this target. This makes diversity metrics,
which quantify the speed at which
error probability is driven to zero with the signal-to-noise ratio, beside the point.

\item Wireless channels in modern systems generally exhibit a
notable amount of time and frequency selectivity, which is naturally converted into diversity benefits through coding and interleaving.
This renders transmit antenna diversity unnecessary.

\item Block error probability is the relevant measure of reliability.
Since the channel codes featured in contemporary systems allow for operation close to
information-theoretic limits, such block error probability is well approximated by the mutual information outages.
Although uncoded error probability is often quantified, this is only an
indirect performance measure, and incorrect conclusions
are sometimes reached by considering only uncoded performance.
\end{itemize}

It is also imperative to recognize that the notion of diversity is indelibly associated with channel uncertainty.
If the transmitter has instantaneous CSI (channel-state information), then it can match its transmission to the channel in such a way that the error probability
depends only on the noise. Diversity techniques, which aim precisely at mitigating the effects of channel uncertainty, are then beside the point.
Although perhaps evident, this point is often neglected. In some models traditionally used to evaluate diversity techniques, for instance,
the channel fades very slowly yet there is no transmitter adaptation.
As we shall see, these models do not reflect the operating conditions of most current systems.

\subsection{Time/Frequency Diversity v. Antenna Diversity}

Perhaps the simplest manifestation of the efficacy of diversity in the aforementioned traditional
settings is receive antenna combining:
if two receive antennas are sufficiently spaced, the same
signal is received over independently faded paths. Even with simple
selection combining, this squares the probability of error; optimal MRC
(maximum-ratio combining) performs even better.

Based upon the specifics of receive antenna combining, it may appear that multiple, independently faded copies of
the same signal are required to mitigate fading. Although this is an accurate description of receive
combining, it is an overly stringent requirement in general. This point is clearly illustrated if one considers a
frequency-selective channel. One simple but na\"{\i}ve method of mitigating fading in such a channel is to repeat the same signal on two sufficiently
spaced frequency channels. Unlike receive combining, this technique doubles
the number of symbols transmitted and therefore the necessary bandwidth.
Is repetition, which seems inefficient, the only way to
take advantage of frequency diversity? It is not---if coding is taken into consideration.
By applying a channel code to a sequence of information bits, the same benefit is gained by transmitting
different portions of the coded block over different frequency channels. No repetition is necessary;
rather, information bits are coded and interleaved,
and then the first half of the coded block is
transmitted on the first frequency and the other half on the second
frequency. The information bits can be correctly
decoded as long as both frequencies are not badly faded.
The same principle applies to time selectivity: instead of repeating the same
signal at different time instants, transmit a coded and interleaved
block over an appropriate time period \cite{TseViswanathBook}.\footnote{When explaining the
exploitation of selectivity through coding and interleaving, it is
important to dispel the misconception that channel coding incurs
a bandwidth penalty.  If the constellation is kept fixed, then
coding does reduce the rate relative to an uncoded system. However, there is
no rate penalty if the constellation size is flexible as in modern systems.
For instance, a system using QPSK
with a rate-$1/2$ binary code and an uncoded BPSK system both have
an information rate of $1$ bit/symbol.  For a reasonably strong code, though, the
coded system will achieve a considerably smaller bit error probability than the
uncoded one. More importantly, the advantage of the coded system in
terms of {\em block} error probability is even larger and this advantage
increases with blocklength: the block error probability of a coded system
decreases with the blocklength whereas, without coding, it actually increases with the blocklength.
As will be emphasized throughout the paper, modern wireless systems cannot be conceived without powerful channel coding.}

\section{Modeling Modern Wireless Systems}
\label{section:modern}

Wireless systems have experienced dramatic changes as they evolved from their initial analog forms to today's
advanced digital formats. Besides MIMO, features of modern systems---that in many cases were completely absent in earlier designs---include:

\begin{itemize}
\item Wideband channelizations and OFDM.
\item Packet switching, complemented with time- and frequency-domain scheduling for low-velocity users.
\item Powerful channel codes \cite{Gallager_LDPC, berrou,Forney_Costello07}.
\item Link adaptation, specifically rate control via variable modulation and coding \cite{andrea2}.
\item ARQ (automatic repeat request) and H-ARQ (hybrid-ARQ) \cite{Costello98}.
\end{itemize}

\noindent
These features have had a major impact on the operational conditions:

\begin{itemize}
\item There is a target block error probability, on the order of
$1\%$, at the output of the decoder. (When H-ARQ is
in place, this target applies at termination.) Link adaptation loops
are tasked with selecting the rate in order to
maintain performance tightly around this
operating point. The rationale for this is two-fold:
\begin{description}
\item{i)} There is little point in spending resources pushing the error probability on the traffic channels much below the error probability on the
control plane, which, by its very nature (short messages and tight latency
requirements), cannot be made arbitrarily small.
\item{ii)} Lower error probabilities often do not improve end-to-end performance:
in some applications (e.g., voice) there is simply no perceivable improvement in
the user experience while, in others (e.g., data communication requiring
very high reliability), it is more cost effective to let
the upper protocol layers handle the losses.
\end{description}
\item The fading of low-velocity users can be tracked
and fed back to the transmitter thereby allowing for link adaptation to the
supportable rate, scheduling on
favorable time/frequency locations, and possibly %transmitter
beamforming and precoding.

\item The channels of high-velocity users vary too quickly
in time to allow for feedback of CSI or even of
the supportable rate. Thus, the signals of such users are
dispersed over the entire available bandwidth thereby taking advantage of
extensive frequency selectivity. In addition, time selectivity is naturally available because of the high velocity.

\end{itemize}

The above points evidence the disparity between the low- and high-velocity regimes
and hence, in order to organize the discussion, it is necessary to
distinguish between them.

\section{The Low-Velocity Regime}

At low velocities, timely feedback regarding the current state of the channel becomes feasible.
This fundamentally changes the nature of the communication problem: all uncertainty is removed except for the noise.
With powerful coding handling that remaining uncertainly, outages are essentially eliminated.\footnote{The rate supported by the channel
may be essentially zero at some time/frequency points, but with proper link adaptation this does not constitute an outage
in the sense that no data is lost (cf. Eq. \ref{eq-outage}).}
Transmit diversity techniques, whose goal is precisely to reduce outages, become inappropriate.
Rate maximization becomes the overriding transmission design principle, and
the optimum strategy in this known-channel setting is spatial waterfilling \cite{telatar}.

Although the above consideration posited perfect CSI at the transmitter, it also extends to imperfect-CSI settings (caused by
limited rate and/or delay in the link adaptation loop).
At a minimum, the supportable rate can be fed back; this still removes outages.
Additional CSI feedback enables adaptive techniques such as
scheduling, power control, beamforming and precoding \cite{JSAClimitedfeedback}.\footnote{Feedback mechanisms are sometimes studied under the assumption that
they convey information regarding the transmit
strategy, e.g., which beamformer or precoder to use, but not regarding
rate selection, in which case outages still occur.
This, however, is not well aligned with modern system designs in which rate control is paramount.}

In multiuser settings, furthermore, CSI feedback is collected from
many users and time- and frequency-domain scheduling offers
additional degrees of freedom. In this case, transmit diversity techniques
can actually be detrimental because they harden the possible
transmission rates to different users thereby reducing potential multiuser
scheduling gains \cite{Hochwald_Hardening04, Viswanathan_Tcom04}.

These conclusions apply almost universally to indoor systems, which conform to
this low-velocity regime, as long as their medium-access control features
the necessary functionalities.
In outdoor systems, they apply to stationary and pedestrian users.

\section{The High-Velocity Regime}
\label{mainsection}

Having established that diversity is not an appropriate perspective in the low-velocity regime, we henceforth focus exclusively on the high-velocity regime.
This is the regime of interest for vehicular users in outdoor systems.
At high velocities, the fading (and therefore the time-varying
mutual information) is too rapid to be tracked. The link adaptation
loops can therefore only match the rate to the average channel conditions.
The scheduler, likewise, can only respond to average conditions and thus it is not possible to
transmit only to users with favorable
instantaneous channels; we thus need not
distinguish between single-user and multiuser settings.

\subsection{Channel Model and Performance Metrics}
\label{sec-prelim}

Let $\nT$ and $\nR$ denote, respectively, the number of transmit and receive antennas.
Assuming that OFDM (orthogonal frequency division multiplexing), the prevalent signalling technique in contemporary systems,
is used to decompose a possibly frequency-selective channel into $N$ parallel, non-interfering tones, the received
signal on the $i$th tone is
\be
\label{LeoMessi}
{\bf y}_i = \Hm_i {\bf x}_i + {\bf n}_i
\ee
where $\Hm_i$ is the $\nR \times \nT$ channel matrix on that tone, ${\bf y}_i$ is the
$\nR \times 1$ received signal, ${\bf n}_i$ is the $\nR \times 1$
thermal noise, IID circularly symmetric complex Gaussian
with unit variance, and ${\bf x}_i$ is the $\nT
\times 1$ transmitted signal subject to a power constraint $\SNR$,
i.e., $E[ || {\bf x}_i ||^2] \leq \SNR$. The
receiver has perfect knowledge of the $N$ channel matrices (the
joint distribution of which is specified later).

For a particular realization of $\Hm_1, \ldots , \Hm_N$, the average mutual information thereon is
\be
\label{chile}
\mathcal{I}(\SNR) = \frac{1}{N} \sum_{i=1}^N I({\bf x}_i; {\bf y}_i).
\ee
This quantity is in bits per (complex) modulation symbol, and thus it represents
spectral efficiency in bits/s/Hz under the standard assumption of one symbol/s/Hz.
 The mutual information on each tone is determined by the chosen signal
distribution. If the signals are IID complex Gaussian\footnote{Actual systems use discrete constellations, for which counterparts to (\ref{eq-logdet}) exist in integral form \cite{mercuryjournal}.
As long as the cardinality of the constellation is large enough relative to the $\SNR$, the gap between the actual mutual information and (\ref{eq-logdet}) is small and inconsequential
to our conclusions.}
with
$E [{\bf x}_i {\bf x}_i^\dagger ] = \frac{\SNR}{\nT} {\bf I}$,
then
\be \label{eq-logdet}
I({\bf x}_i; {\bf y}_i) =  \log_2 \det \left( {\bf I} + \frac{\SNR}{\nT} \, \Hm_i
\Hm_i^\dagger  \right).
\ee
Since approaching this mutual information may entail high complexity, simpler MIMO strategies
with different (lower) mutual informations are often used.
Expressions for these are given later in this Section.

Once a transmission strategy has been specified, the corresponding
outage probability for rate $R$ (bits/s/Hz) is then
\be
\label{eq-outage}
P_{\sf out}(\SNR,R) = \mathrm{Pr} \{ \mathcal{I} (\SNR) < R \}.
\ee
With suitably powerful codes, the error probability when not in outage is very small and
 the outage probability is an accurate approximation for
the actual block error probability \cite{ozarow}--\nocite{CaTaBi}\cite{PrasadVaranasi_IT06}. We shall therefore use both notions
interchangeably henceforth.

As justified in Section \ref{section:modern}, modern systems operate at a target error probability. Hence,
the primary performance metric is the maximum rate, at each $\SNR$, such that this target is not exceeded, i.e.,
\be
\label{bharti}
R_{\epsilon}(\SNR) = \max_{\zeta} \{ \zeta: P_{\sf out}(\SNR,\zeta) \leq \epsilon \}
\ee
where $\epsilon$ is the target.

\subsection{The Diversity-Multiplexing Tradeoff}

A clear tradeoff exists between rate,
outage and $\SNR$.  Traditionally, notions of diversity order study
the speed at which error probability decreases (polynomially) as
$\SNR$ is taken to infinity while $R$ is kept fixed as in (\ref{eq-outage}).
Although meaningful in early wireless systems, where $R$ was
indeed fixed, this is not particularly indicative of contemporary
systems in which $R$ is increased with $\SNR$.

An alternative formulation was introduced in \cite{divmul}, where
$R$ increases with $\SNR$ according to some function $R=f(\SNR)$. % while $P_{\sf out}(\SNR)$ decreases with $\SNR$.
The multiplexing gain is defined as
\bea
\label{equation:r}
r = \lim_{{\sf \scriptscriptstyle SNR} \rightarrow \infty} \frac{f(\SNR)}{\log \SNR},
\eea
which is the asymptotic slope of the rate-$\SNR$ curve
in bits/s/Hz/ (3 dB), while the diversity order is defined as
\be
\label{equation:d}
d = - \lim_{{\sf \scriptscriptstyle SNR} \rightarrow \infty} \frac{\log P_{\sf out}(\SNR,f(\SNR))}{\log \SNR}.
\ee
Given a number of
transmit and receive antennas, diversity and multiplexing are
conflicting objectives as succinctly captured by the DMT
(diversity-multiplexing tradeoff) \cite{divmul}. Formulated for a
quasi-static channel model where each coded block is subject to a
single realization of the fading process, the DMT specifies that,
with $\nT$ transmit and $\nR$ receive antennas, $\min(\nT,\nR)+1$
distinct DMT points are feasible, each corresponding to a
multiplexing gain $0\leq r \leq \min(\nT,\nR)$ and a diversity order
\be
\label{equation:dmt}
d(r)=(\nT-r)(\nR-r) .
\ee
Stated simply, if the rate is increased with $\SNR$ as $r \log \SNR$ then
the outage can decrease no faster than $\SNR^{-(\nT-r)(\nR-r)}$.
This is the optimum DMT; then, each specific transmit-receive architecture is associated with a DMT that may or may not achieve this optimum.

Note that, in (\ref{equation:dmt}) and throughout the paper, $d$
quantifies only the \emph{antenna} diversity order as per the asymptotic
definition in (\ref{equation:d}). If the coded block spans several
fading realizations, then this additional time/frequency selectivity leads to
larger diversity orders but does not increase the maximum value of $r$ \cite{divmul,CoronelBolcskei07}.

A multiplexing gain $r=0$ signifies a rate that does not increase
(polynomially) with the $\SNR$ while $d=0$ indicates an outage
probability that does not decrease (polynomially) with the $\SNR$.

Although the DMT is a powerful tool, it has
clear limitations that stem from the fact that the diversity order and the multiplexing gain are
only proxies for performance measures of real interest (error probability and rate, respectively).
The asymptotic nature of the definitions of $r$ and $d$ naturally
restricts the validity of the DMT insights to the high-power regime.\footnote{Any feature whose effect is non-polynomial in the $\SNR$ is immaterial in terms
of the DMT.
Non-asymptotic DMT formulations, valid for arbitrary $\SNR$, have been put forth but
they lack the simplicity and generality of (\ref{equation:dmt}) \cite{Narasimhan06,AzarianElGamal07}.}
Even in that regime, the diversity order does not suffice to determine the error probability at a given $\SNR$.
It simply quantifies the speed at which the error probability falls with the $\SNR$.
Similarly, the multiplexing gain does not suffice to determine the rate, but it only quantifies how the rate grows with the $\SNR$.

The quantity of interest $R_{\epsilon}(\SNR)$ introduced in (\ref{bharti}) corresponds to the $d=0$ DMT point.
From the DMT, all we can infer about it is the value of the asymptotic slope
\be
\lim_{\SNR \rightarrow \infty}  \frac{R_{\epsilon}(\SNR)}{\log \SNR}
\ee
which can, at most, equal $\min(\nT,\nR)$. Certain architectures achieve this maximum, while others fall short of it.
The traditional notion of diversity, in turn, provides no information about $R_{\epsilon}$ because it is defined for some fixed rate.

\subsection{Diversity v. Multiplexing in Modern Systems}

In this high-velocity scenario, frequency-flat analyses are likely to indicate that dramatic reductions in outage probability
can be had by increasing $d$. On these grounds, transmission strategies that operate efficiently at the full-diversity
DMT point have been developed.
The value of these strategies for modern wireless systems, however, is questionable because:

\begin{enumerate}
\item The outage need not be reduced below the target error probability.
\item Diversity is plentiful already:
\begin{description}
\item{i)} By the same token that the fading is too rapid to be tracked, it offers time selectivity.
\item{ii)} Since, in this regime, modern systems distribute the signals over large swaths of bandwidth,
there tends to be abundant frequency selectivity.
\end{description}
\end{enumerate}

Within the DMT framework, a fixed outage probability corresponds to $d=0$, i.e., to the full multiplexing gain achievable
by the architecture at hand. Thus, as recognized in \cite{Barry}, the $R_\epsilon$-maximizing architectures for $\SNR \rightarrow \infty$ are those that can attain the maximum multiplexing
gain $r=\min(\nT,\nR)$.
Due to the nature of the DMT, however, this holds asymptotically in the $\SNR$.
The extent to which it holds for $\SNR$ values of interest in a selective channel can only be determined
through a more detailed (non-asymptotic) study.
To shed light on this point, a case study is presented next.

\subsection{Case Study: A Contemporary MIMO-OFDM System}

Let us consider the exemplary system described in Table~\ref{table:parameters}, which is
loosely based on the 3GPP LTE design \cite{LTEbook}. (With only slight modifications, this
system could be made to conform with 3GPP2 UMB or with IEEE 802.16 WiMAX.)
Every feature relevant to the discussion at hand is modeled:

\begin{table}
\centering \caption{MIMO-OFDM System Parameters} \label{table:parameters}
\begin{tabular}{|l|l|} \hline\hline
Tone spacing & $15$ kHz \\ \hline OFDM Symbol duration & 71.5 $\mu$s \\ \hline Bandwidth & $10$ MHz ($600$ tones, excluding guards) \\ \hline Resource
block & $12$ tones over $1$ ms ($168$ symbols) \\ \hline H-ARQ & Incremental redundancy \\ \hline H-ARQ round spacing & $6$ ms \\ \hline Max. number H-ARQ
rounds & 6 \\ \hline Power delay profile & 12-ray TU \\ \hline Doppler spectrum & Clarke-Jakes \\ \hline Max. Doppler frequency & $185$ Hz \\ \hline
Antenna correlation & None \\ \hline\hline
\end{tabular}
\end{table}

\begin{itemize}
\item A basic resource block spans $12$ OFDM tones over $1$ ms. Since $1$ ms corresponds to $14$ OFDM symbols, a
resource block consists of $168$ symbols. In the high-velocity
regime being considered, the $12$ tones are interspersed uniformly over
$10$ MHz of bandwidth. There are $600$ usable tones on that
bandwidth, guards excluded, and hence every $50$th tone is
allocated to the user at hand while the rest are available for other
users.\footnote{For low velocity users, in contrast, the $12$ tones in a resource block are contiguous so that their fading can be
efficiently described and fed back for link adaptation and scheduling purposes as discussed in Section \ref{section:modern}.}

\item Every coded block spans up to $6$ H-ARQ transmission rounds, each corresponding to a basic resource block,
with successive rounds spaced by $6$ ms for a maximum temporal span of $31$ ms. (This is an acceptable delay for most applications,
including Voice-over-IP.)
The H-ARQ process terminates as soon as decoding is possible. An error is declared if decoding is not possible after
$6$ rounds.

\item The channel exhibits continuous Rayleigh fading with a Clarke-Jakes spectrum and a $180$-Hz maximum Doppler frequency.
(This could correspond, for example, to a speed of $100$ Km/h at $2$
GHz.) The power delay profile is given by the $12$-ray TU (typical
urban) channel detailed in Table~\ref{table:pdp}. The r.m.s. delay
spread equals $1$ $\mu$s.

\item The antennas are uncorrelated to underscore the roles of both diversity and multiplexing.
Some comments on antenna correlation are put forth later in the section.

\end{itemize}

\begin{table}
\centering \caption{TU power delay profile} \label{table:pdp}
\begin{tabular}{|l|l|} \hline\hline
Delay ($\mu$s) & Power (dB) \\ \hline\hline 0 & -4 \\ \hline 0.1 & -3 \\ \hline 0.3 & 0 \\ \hline 0.5 & -2.6 \\ \hline 0.8 & -3 \\ \hline 1.1 & -5 \\
\hline 1.3 & -7 \\ \hline 1.7 & -5 \\ \hline 2.3 & -6.5 \\ \hline 3.1 & -8.6 \\ \hline 3.2 & -11 \\ \hline 5 & -10 \\ \hline\hline
\end{tabular}
\end{table}

The impulse response describing each of the $\nT \nR$ entries of the channel matrix is
\be
h(t,\tau) = \sum_{j=1}^{12} \sqrt{\alpha_j} c_j(t) \delta(t - \tau_j)
\ee
where the delays $\{ \tau_j \}_{j=1}^{12}$ and the powers $\{ \alpha_j \}_{j=1}^{12}$ are specified in
Table~\ref{table:pdp} and $\{ c_j(t) \}_{j=1}^{12}$ are
independent complex Gaussian processes with a Clarke-Jakes spectrum.
Although time-varying, the channel is suitably constant for the duration of an OFDM symbol
such that it is meaningful to consider its frequency response as in (\ref{LeoMessi}).

\begin{figure}
\centerline{\psfig{figure=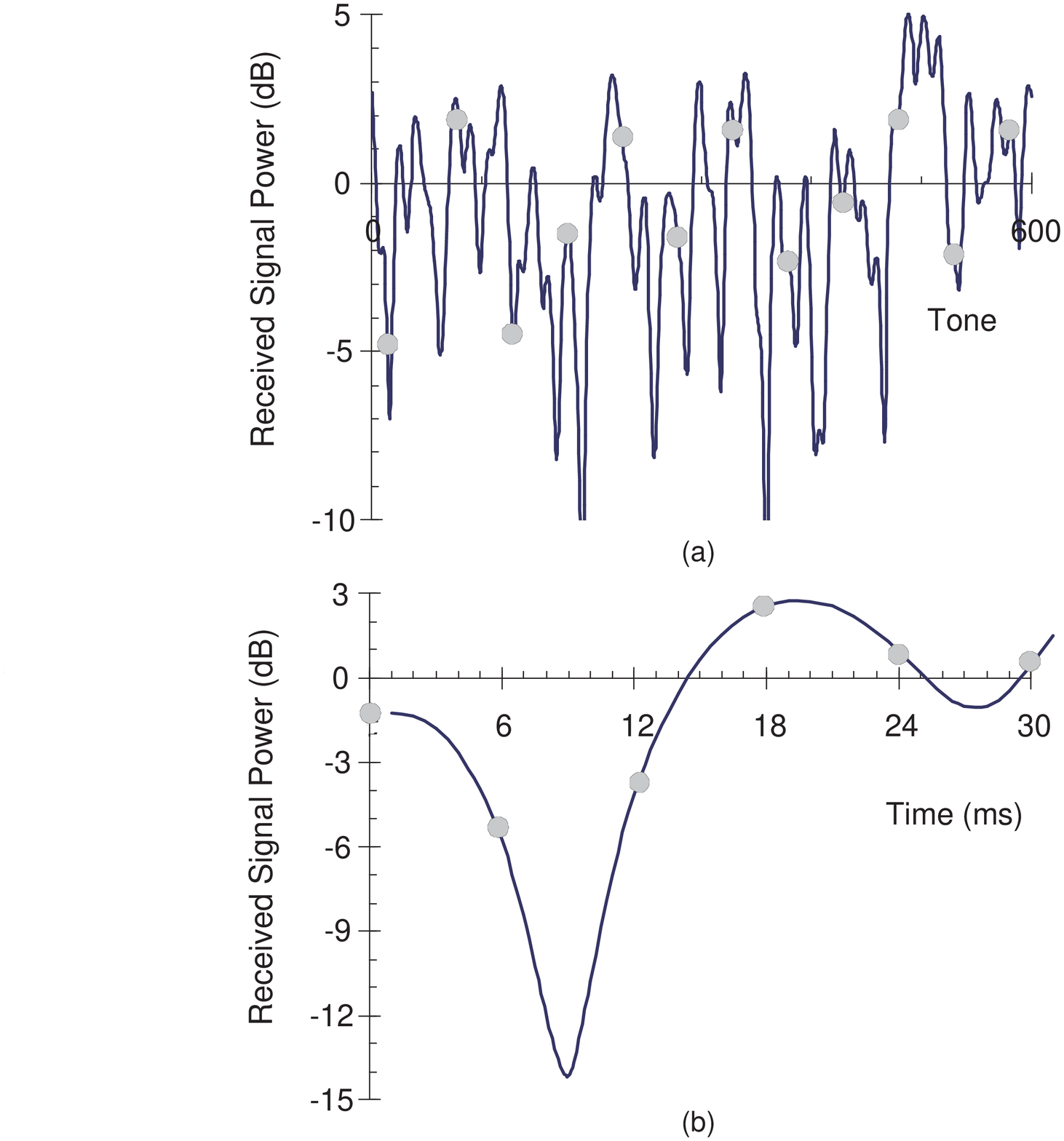,width=8.0in}} \caption{(a) TU channel fading realization over $600$ tones. The circles indicate the locations of the
$12$ tones that map to a given resource block. (b) TU channel fading realization, for a given tone, over $30$ ms. The circles indicate the locations of
the $6$ H-ARQ rounds.} \label{fig:channel}
\end{figure}

The variability of the channel response over the
multiple tones and H-ARQ rounds of a coded block is illustrated in Fig.~\ref{fig:channel}.
Note the very high degree of frequency selectivity and how the channel decorrelates during the $6$ ms separating H-ARQ rounds.

Without H-ARQ, rate and outage are related as per (\ref{eq-outage}).
With H-ARQ, on the other hand, the length of each
coded block becomes variable. With IR (incremental redundancy)
specifically, mutual information is accumulated over successive
H-ARQ rounds \cite{Caire_Tuninetti}. If we let
$\mathcal{M}_k(\SNR)$ denote the mutual information after $k$
rounds, then the number of rounds needed to decode a particular
block is the smallest integer $K$ such that
\be
\mathcal{M}_K(\SNR)> 6 \, R_{\epsilon}(\SNR)
\ee
where %${\mathcal R}_1$ indicates the rate of the initial round and
$K \leq 6$. A one-bit notification of
success/failure is fed back after the receiver attempts to decode
following each H-ARQ round. An outage is declared if $\mathcal{M}_6(\SNR) \leq 6 \, R_{\epsilon}(\SNR)$ and the effective rate
(long-term average transmitted rate) is
\be
{\mathcal R}_{\epsilon}(\SNR) = \frac{6 \, R_{\epsilon}(\SNR)}{E[K]}
\label{eq-arq_rate}
\ee
The initial rate is selected such that the outage at H-ARQ termination is precisely $\epsilon=1\%$.
This corresponds to choosing an initial rate of $6 \, R_{\epsilon}$ where $R_{\epsilon}$ corresponds to the quantity of interest defined in (\ref{bharti})
with the mutual information averaged over the $168$ symbols within each H-ARQ round and then summed across the $6$ rounds.

In order to contrast the benefits of transmit diversity and spatial multiplexing,
we shall evaluate two representative transmission techniques:

\begin{itemize}
\item A transmit diversity strategy that converts the MIMO channel into an effective
scalar channel with signal-to-noise ratio
\be
\label{laureta}
\frac{\SNR}{\nT} \, \Tr \left\{ \Hm_i(k) \Hm_i^\dagger(k) \right\}
\ee
where $\Hm_i(k)$ denotes the channel for the $i$th symbol on the $k$th H-ARQ round.
By applying a strong outer code to this effective scalar channel,
the mutual information after $k$ rounds is, at most \cite{divmul}
\be \label{eq-diversity_mi}
\mathcal{M}_k(\SNR) = \sum_{\ell = 1}^k \frac{1}{168}  \sum_{i=1}^{168}  \log \left( 1 + \frac{\SNR}{\nT} \, \Tr \left\{ \Hm_i(\ell) \Hm_i^\dagger(\ell) \right\} \right)
\ee
Transmit diversity strategies provide
full diversity order with reduced complexity, but their
multiplexing gain cannot exceed $r=1$, i.e., one information symbol for every vector ${\bf x}_i$ in (\ref{LeoMessi}).
Note that, when $\nT=2$, (\ref{eq-diversity_mi}) is achieved by Alamouti transmission \cite{alamouti}.

\item A basic MMSE-SIC spatial multiplexing strategy where a separate coded signal is transmitted from each antenna,
all of them at the same rate \cite{jerry3}.\footnote{Separate rate control of each coded signal based on instantaneous channel conditions can make this strategy
optimal in terms of outage \cite{PARC}, but this is infeasible in this high-velocity regime.}
The receiver attempts to decode the signal transmitted from the first antenna. An MMSE filter is applied to whiten the interference from the
other signals, which means that the first signal experiences a signal-to-noise ratio
\be
{\bf h}^\dagger_{i,1}(k) \left( \Hm_{i,1}(k) \Hm^\dagger_{i,1}(k) + \frac{\nT}{\SNR} {\bf I} \right)^{-1} {\bf h}_{i,1}(k).
\ee
during the $k$th H-ARQ round. If successful, the
effect of the first signal is subtracted from the received samples and decoding
of the second signal is attempted, and so forth.  No optimistic
assumption regarding error propagation is made: an outage is declared if any of the $\nT$ coded signals cannot support the transmitted rate.
The aggregate mutual information over the $\nT$ antennas after $k$ H-ARQ rounds is
$$
\hspace{-5mm}
\mathcal{M}_k(\SNR) = \nT \min_{m=1,\cdots,\nT} \left \{  \sum_{\ell = 1}^k  \frac{1}{168} \sum_{i=1}^{168} \log \left( 1 + {\bf h}^\dagger_{i,m}(\ell) \left( \Hm_{i,m}(\ell) \Hm^\dagger_{i,m}(\ell) +\frac{\nT}{\SNR} {\bf I} \right)^{-1} {\bf h}_{i,m}(\ell) \right) \right \}
$$
where ${\bf h}_{i,m}(\ell)$ is the $m$th column of $\Hm_i(\ell)$ and $\Hm_{i,m}(\ell) = [{\bf h}_{i,m+1}(\ell) {\bf h}_{i,m+2}(\ell) \cdots {\bf h}_{i,\nT}(\ell)]$.
While deficient in terms of diversity order, this strategy yields full multiplexing gain, $r = \min(\nT,\nR)$, when $d=0$.
This MMSE-SIC structure is representative of the single-user MIMO mode in LTE \cite{LTEbook}.
\end{itemize}

\begin{figure}
\centerline{\psfig{figure=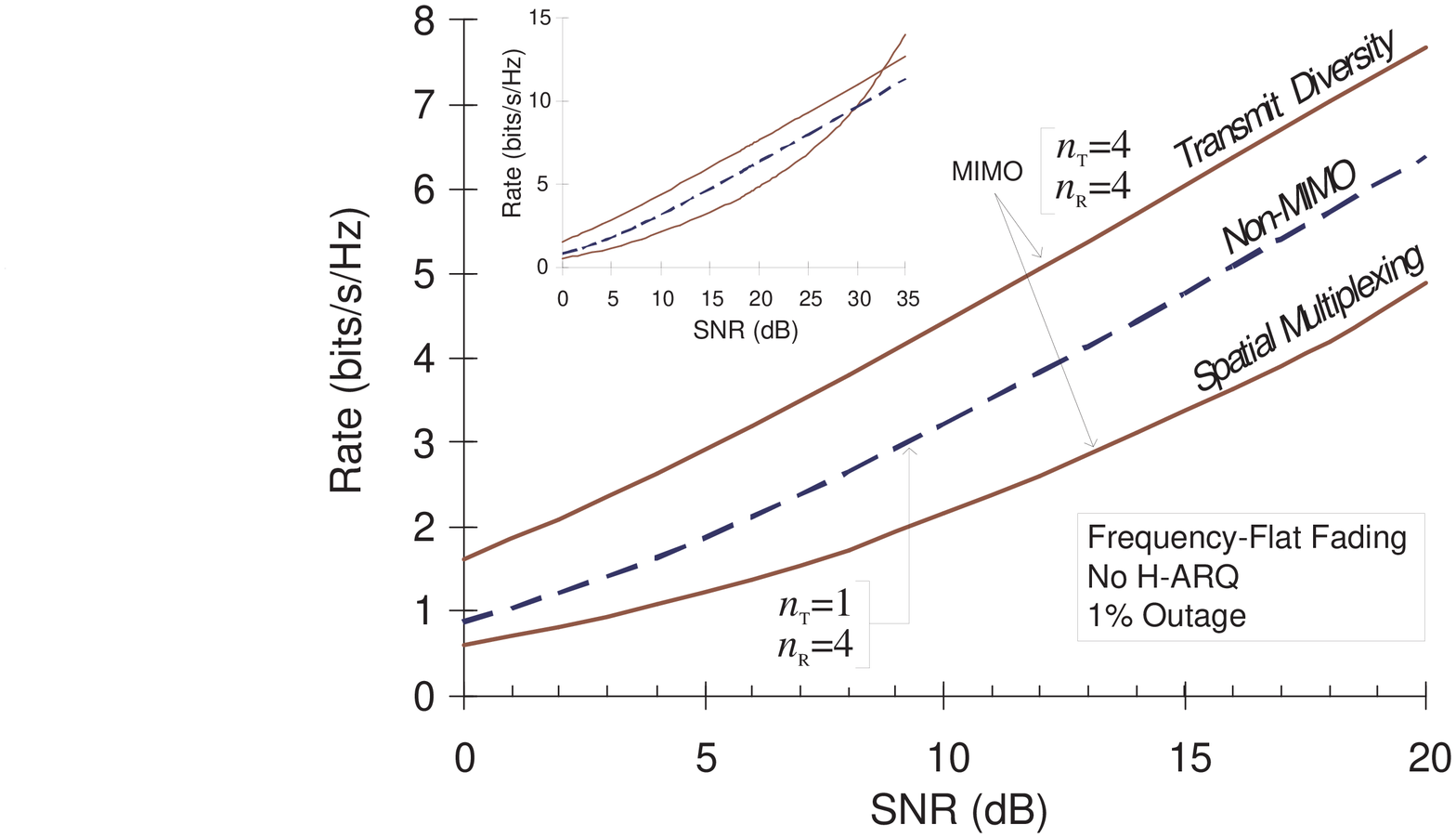,width=8.0in}}
%\centering
%\includegraphics{MI2old.eps}
\caption{Main plot: MMSE-SIC spatial multiplexing v. transmit diversity with $\nT=\nR=4$ in a frequency-flat Rayleigh-faded channel with no H-ARQ. Also
shown is the non-MIMO reference ($\nT=1$, $\nR=4$). Inset: Same curves over a wider $\SNR$ range.} \label{fig:MI2_old}
\end{figure}

Let $\nT=\nR=4$, the high-end configuration for LTE, and consider first a simplistic model
where the fading is frequency-flat and there is no H-ARQ. Every coded block is
therefore subject to a single realization of the Rayleigh fading process.
Under such model, the spectral efficiencies achievable with $1\%$ outage, $\mathcal{R}_{0.01}(\SNR)$, are
compared in Fig.~\ref{fig:MI2_old}
alongside the corresponding efficiency for the non-MIMO reference ($\nT=1$, $\nR=4$).
Transmit diversity is uniformly superior to
spatial multiplexing in the $\SNR$ range of interest.
In fact, spatial multiplexing results in a loss with respect to non-MIMO transmission
with the same number of receive antennas.
The curves eventually cross, as the DMT predicts
and the inset in Fig. \ref{fig:MI2_old} confirms (the asymptotic slope of spatial multiplexing is $r=4$ bits/s/Hz/(3 dB) while $r=1$ for transmit diversity and for non-MIMO), but this crossover does not
occur until beyond $30$ dB.

\begin{figure}
\centerline{\psfig{figure=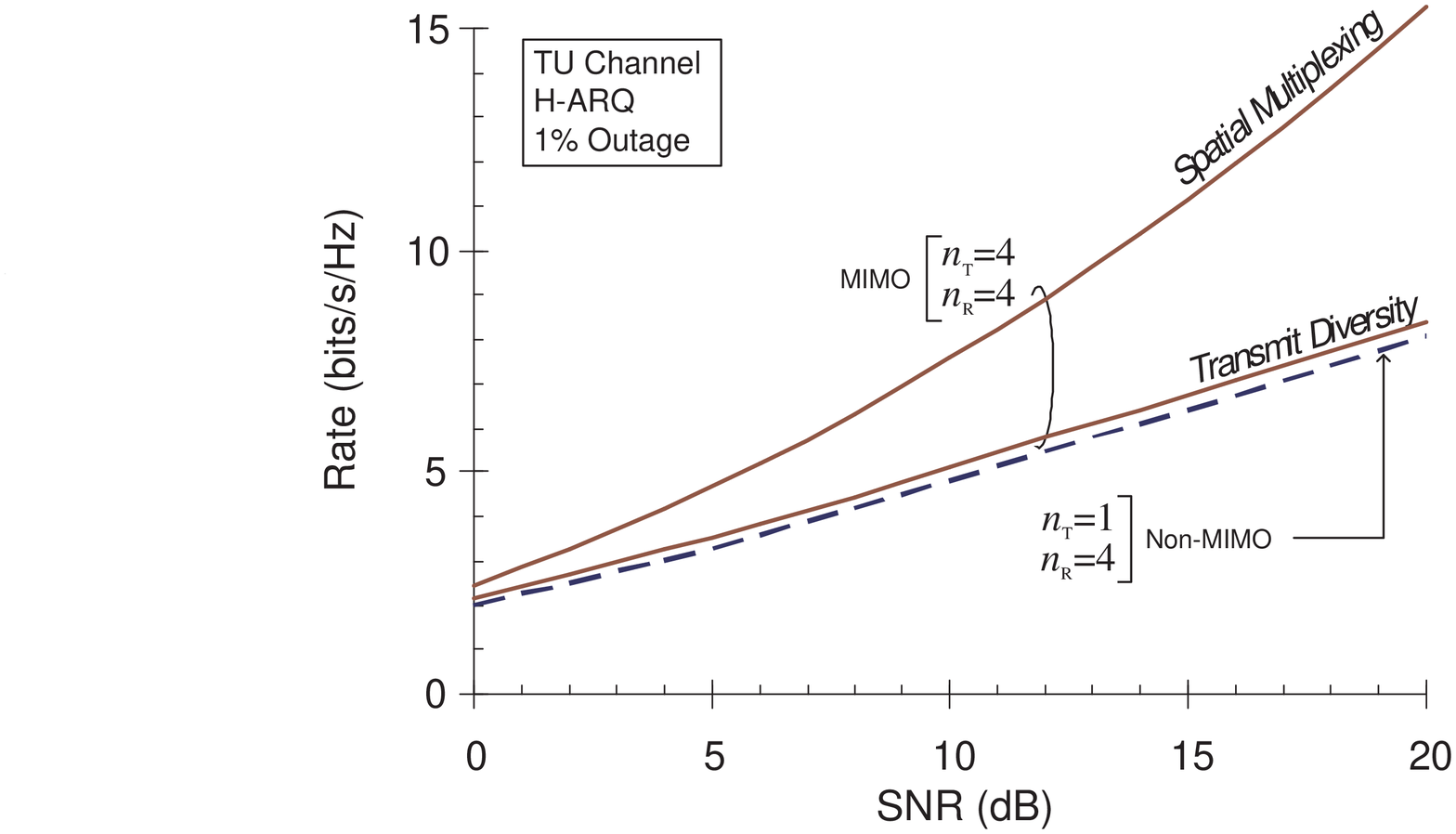,width=8.0in}} \caption{MMSE-SIC spatial multiplexing v. transmit diversity with $\nT=\nR=4$ in the channel described
in Tables~\ref{table:parameters}--\ref{table:pdp}. Also shown is the non-MIMO reference ($\nT=1$, $\nR=4$).} \label{fig:MI2_new}
\end{figure}

Still with $\nT=\nR=4$, consider now the richer model described in Tables~\ref{table:parameters}--\ref{table:pdp}.
The effective mutual information for each block is averaged over tones and symbols and accumulated over H-ARQ rounds.
The corresponding comparison is presented in Fig.~\ref{fig:MI2_new}. In this case, transmit diversity offers a negligible advantage
whereas spatial multiplexing provides ample gains with respect to non-MIMO.

\begin{figure}
\centerline{\psfig{figure=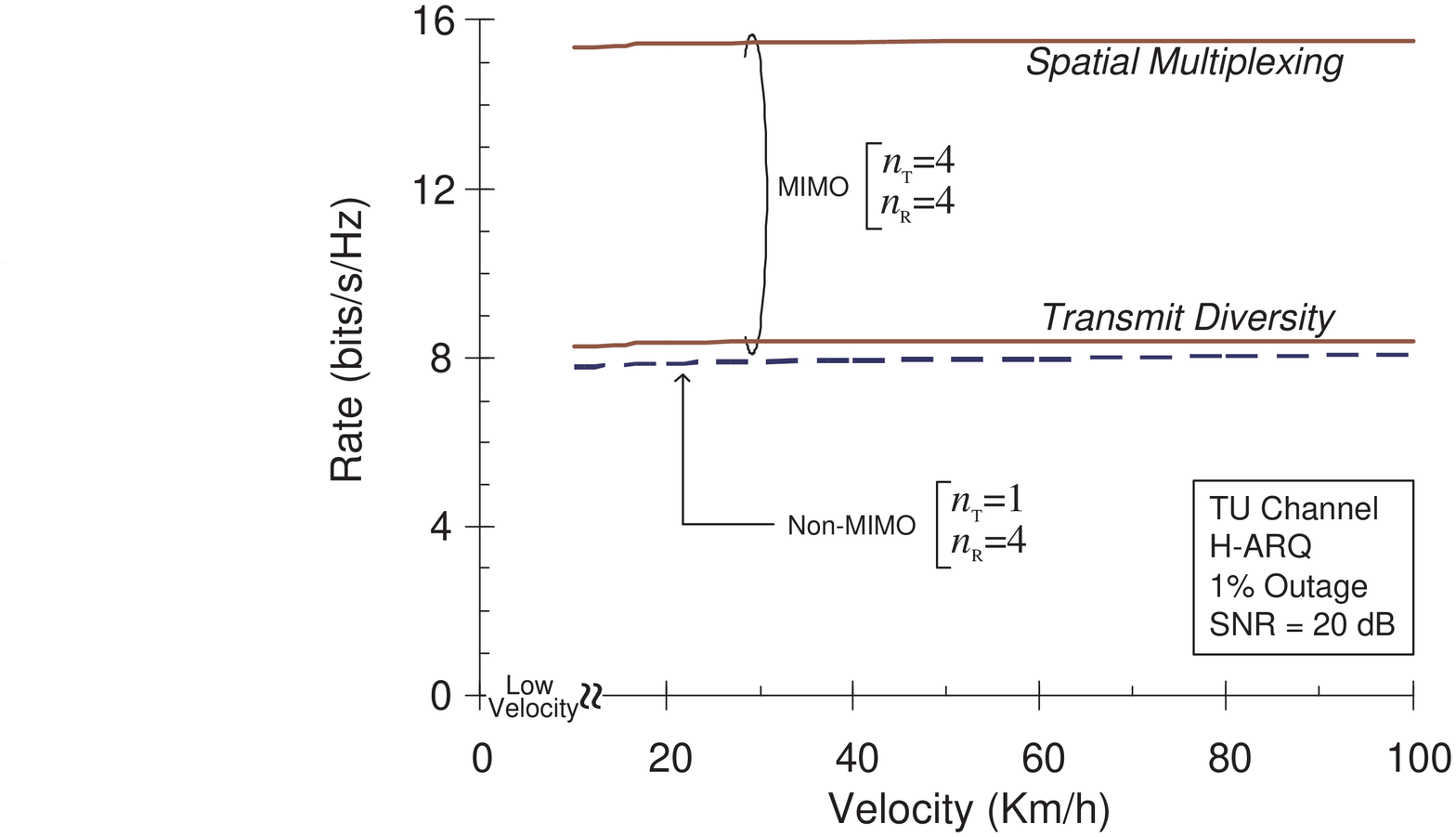,width=8.0in}} \caption{MMSE-SIC spatial multiplexing, transmit diversity and non-MIMO transmission as function of
velocity for the channel described in Tables~\ref{table:parameters}--\ref{table:pdp} at $\SNR = 20$ dB. (Below some point, the system transitions to the
low-velocity regime and thus the curves are no longer meaningful.)} \label{fig:MI_speed}
\end{figure}

The stark contrast between the behaviors observed under the
different models can only be explained by the abundant time and
frequency selectivity neglected by the simple model and actually
present in the system. This renders transmit antenna diversity superfluous,
not only asymptotically but at every $\SNR$. Under the
frequency-flat model, the signal from the first antenna in the
spatial multiplexing transmission does not benefit from any spatial
diversity and thus a low rate must be used so that this signal (and
the subsequent ones) can be decoded with sufficient probability.
Under the richer channel model, however, the first signal reaps diversity from
time/frequency selectivity and thus the lack of spatial
diversity is essentially inconsequential. This behavioral contrast
is, moreover, highly robust. Even if the speed is reduced down to
where the low-velocity regime might start, as in Fig.
\ref{fig:MI_speed}, the behaviors are hardly affected because there
is still significant selectivity. Likewise, the performances are largely
preserved if the bandwidth is diminished significantly below $10$
MHz or the delay spread is reduced below $1$ $\mu$s.

%%%%%%%%%%%%%%%%%%%%%%%%%%%%%%%%%%%%%%%%%%%%%%%%%%%%%%%%%%%%%%%%%%%%%%%%%
\subsection{Ergodic Modeling}

\begin{figure}
\centerline{\psfig{figure=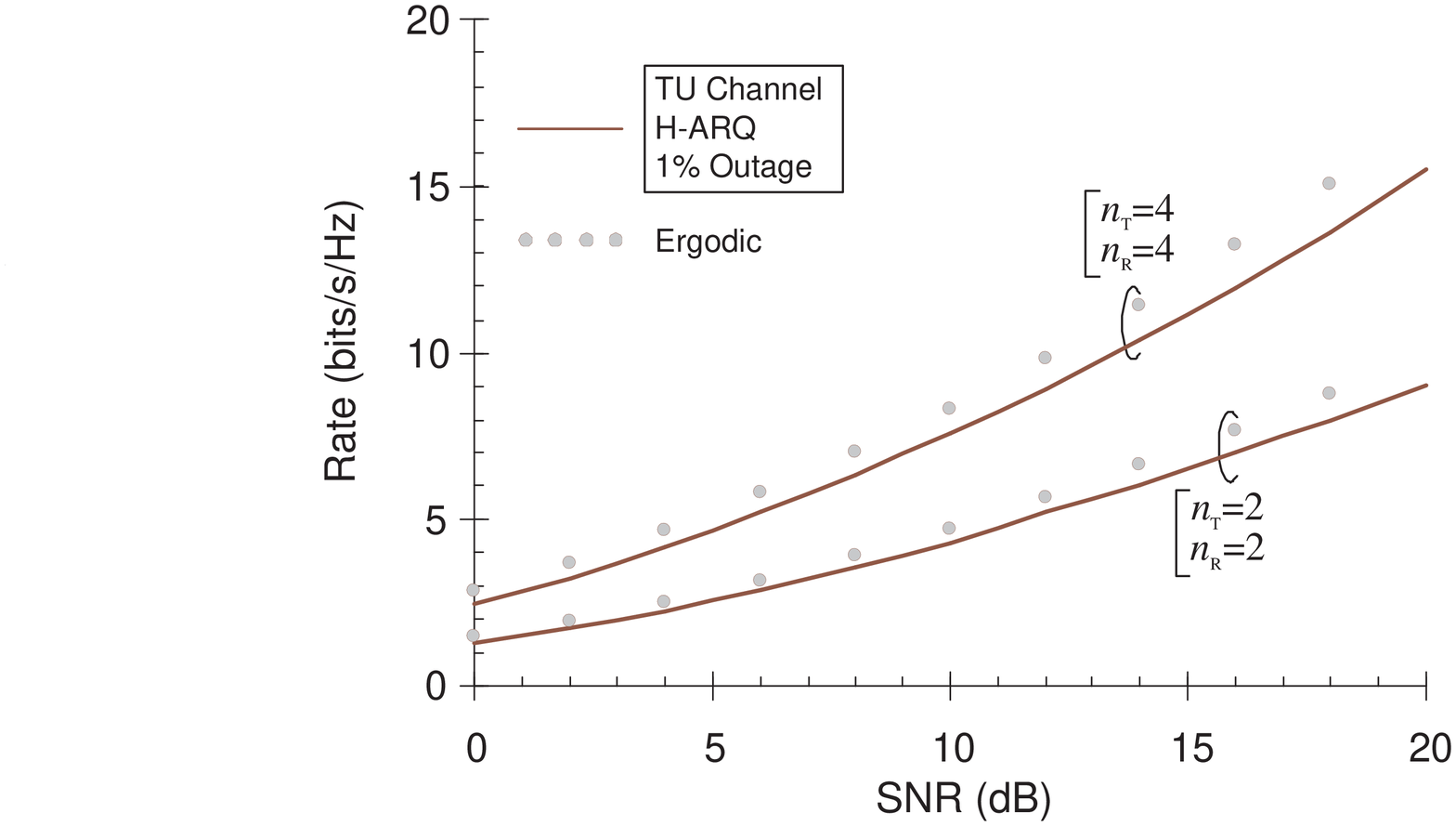,width=8.0in}} \caption{In solid lines, $1\%$-outage rate achievable with MMSE-SIC spatial multiplexing in the channel
described in Tables~\ref{table:parameters}--\ref{table:pdp}. In circles, corresponding ergodic rate for the same numbers of antennas.} \label{fig:MI_erg}
\end{figure}

As it turns out, the time/frequency selectivity in modern systems
is so substantial as
to justify the adoption of an ergodic model altogether. Shown in
Fig.~\ref{fig:MI_erg} is the correspondence between the exact
rates achievable with $1 \%$ outage in the channel described in Tables~\ref{table:parameters}--\ref{table:pdp} and
the respective ergodic rates.

From a computational
standpoint, this match is welcome news because of the fact that convenient
closed forms exist for the rates achievable in an
ergodic Rayleigh-faded channel \cite{shinlee}. Moreover, the optimum
transmission strategies and the impact upon capacity
of more detailed channel features such as antenna correlation, Rice
factors, colored out-of-cell interference, etc, can then be asserted by virtue
of the extensive body of results available for the ergodic setting \cite{poweroffset,ImpactCorr}.

Antenna correlation, for example,
leads to a disparity in the distribution of the spatial eigenmodes that effectively reduces the
spatial multiplexing capability.  Such
effects should, of course, be taken under consideration when determining
the appropriate transmission strategy.

\subsection{Optimal MIMO Detection}

While in the case study we considered the performance of a low complexity but
suboptimal detection scheme for spatial multiplexing, the continual increase
in computational power is now rendering optimal or near-optimal MIMO detection
feasible.  Rather than transmitting separate coded signals from the $\nT$ antennas,
a single one can be interleaved over time, frequency
and the transmit antennas.
At the receiver side, each vector symbol is then fed to a detector that derives soft
estimates of each coded bit---possibly by use of a sphere decoder---to
a standard outer decoder (e.g., message-passing decoder), with subsequent
iterations between the MIMO detector and
the decoder \cite{HochwaldBrink03}. Such techniques, and others such as mutual information lossless codes \cite{PerfectCodes}, can approach
the mutual information in (\ref{eq-logdet}).\footnote{It should be emphasized that
these approaches perform optimal or near-optimal
detection of each MIMO vector symbol, but do not attempt optimal detection
of the outer code. Hence, complexity increases exponentially with
the multiplexing order and the constellation cardinality, but not with the
blocklength.}

In the context of our comparison between transmit diversity and
spatial multiplexing, it is worthwhile to note that the mutual
information in (\ref{eq-logdet}) is greater than that
of transmit diversity for {\em any} channel matrix
$\Hm$. Denoting by $\lambda_\ell$ the $\ell$th eigenvalue of $\Hm
\Hm^\dagger$, \bea \log \det \left( {\bf I} + \frac{\SNR}{\nT} \,
\Hm \Hm^\dagger  \right) &=& \log \left( \prod_{\ell=1}^{\nR}
\left(1 + \frac{\SNR}{\nT} \lambda_\ell \right)
\right) \label{recife1} \\
 &\geq& \log \left( 1 + \frac{\SNR}{\nT} \sum_{\ell=1}^{\nR} \lambda_\ell \right)  \label{recife2} \\
&=&  \log \left( 1 + \frac{\SNR}{\nT} \, \Tr \left\{ \Hm \Hm^\dagger \right\} \right) \label{recife3}
\eea
where (\ref{recife1}) holds because the determinant equals the
product of the eigenvalues, (\ref{recife2}) comes from dropping terms in the product, and (\ref{recife3}) follows from
$\Tr \left\{ \Hm \Hm^\dagger \right\}  = \sum_{\ell=1}^{\nR} \lambda_\ell$.

\begin{figure}
\centerline{\psfig{figure=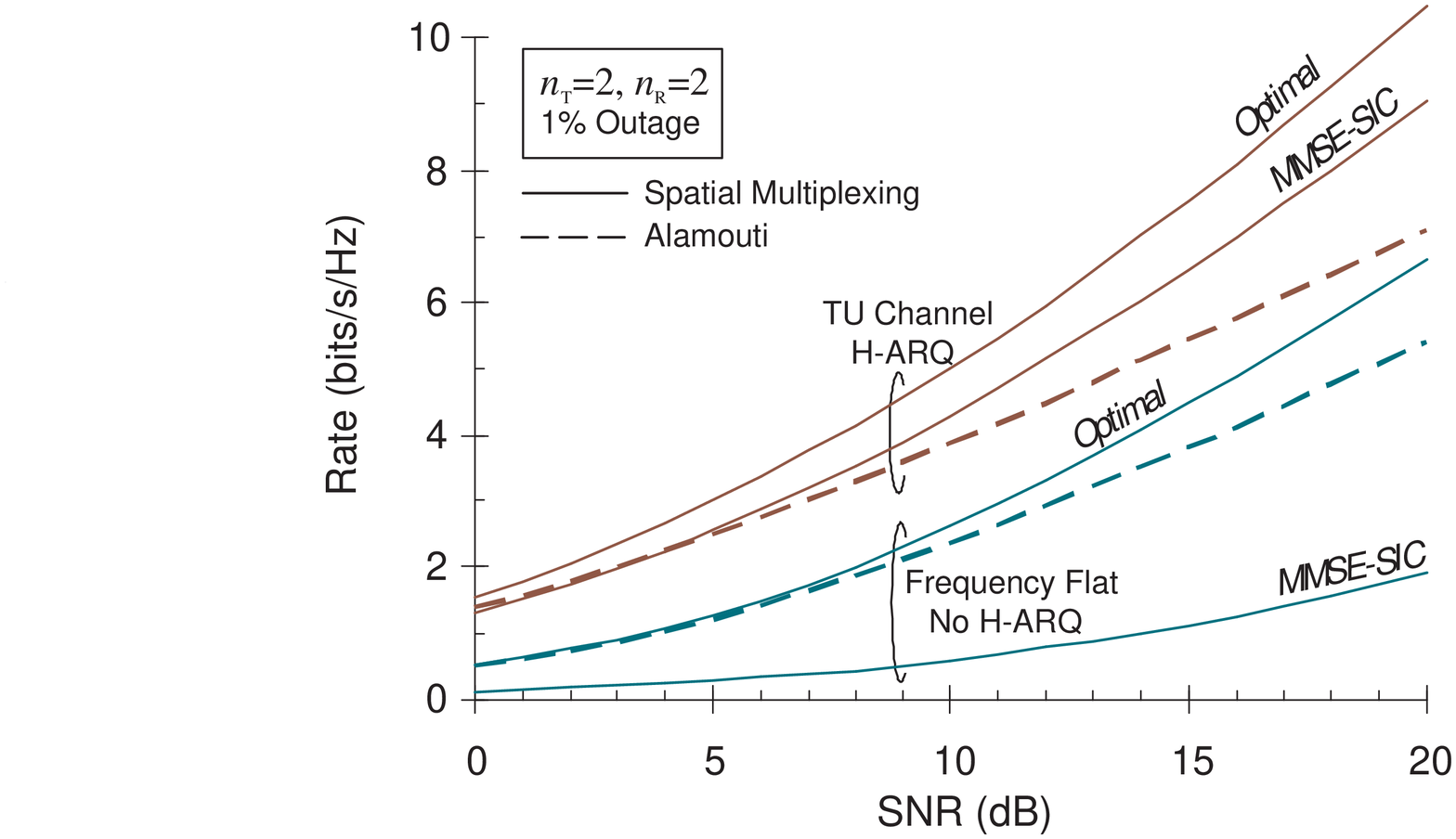,width=8.0in}} \caption{Spectral efficiencies achievable with Alamouti transmission and with spatial multiplexing
(optimal and MMSE-SIC) for $\nT=\nR=2$. The comparisons are shown for both a frequency-flat channel without H-ARQ and for the channel described in
Tables~\ref{table:parameters}--\ref{table:pdp}.} \label{fig:optimal}
\end{figure}

Hence, spatial multiplexing with optimal detection is uniformly superior
to transmit diversity: there truly is no decision to be made between the two
architectures if optimal MIMO detection is an option.
Drawing parallels with the discussion in Section \ref{intro} about the suboptimality of repeating the same signal on two frequency channels versus
transmitting different portions of a coded block thereon, one could equate transmit diversity with the former and the optimum MIMO strategy with the latter.

There is another interesting parallel to our earlier discussion regarding the
importance of channel modeling.  In Fig. \ref{fig:optimal},
the spectral efficiencies of Alamouti transmission and spatial multiplexing (with optimal detection and MMSE-SIC)
are shown for $\nT=\nR=2$, for both the frequency-flat model and the richer model
in Tables~\ref{table:parameters}--\ref{table:pdp}.  Optimal spatial multiplexing
is superior to Alamouti with both models, as per the above derivation, but the
difference is considerably larger when the rich model is used.
Indeed, based upon the frequency-flat model one might
incorrectly conclude that spatial multiplexing provides only a negligible
advantage over Alamouti.  Note also that MMSE-SIC performs well below
Alamouti in the frequency-flat setting, but outperforms it in the rich model.

%%%%%%%%%%%%%%%%%%%%%%%%%%%%%%%%%%%%%%%%%%%%%%%%%%%%%%%%%%%%%%%%%%%%%%%%%
\section{Uncoded Error Probability: A Potentially Misleading Metric}

\begin{figure}
\centerline{\psfig{figure=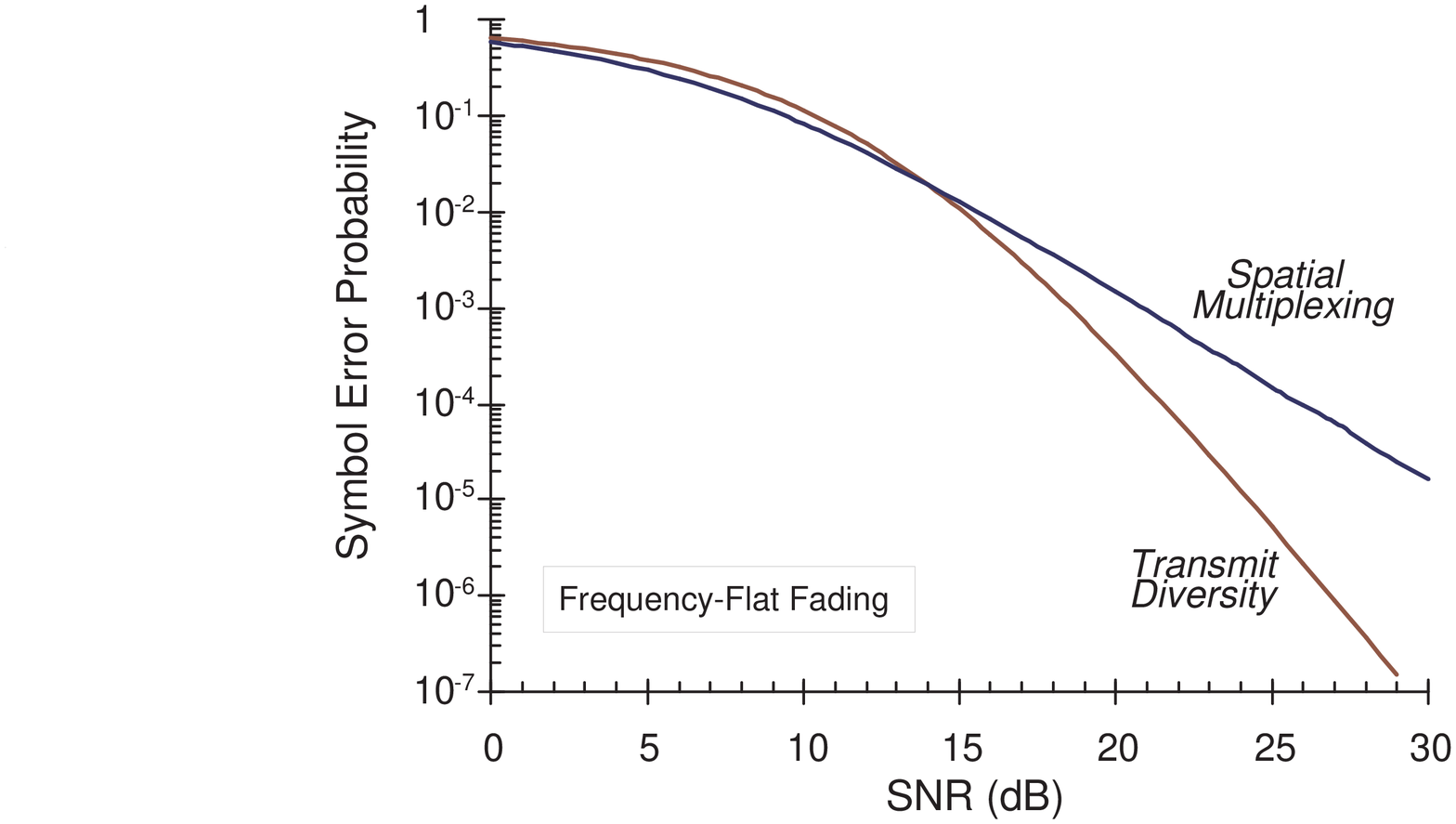,width=8.5in}} \caption{Uncoded symbol error probability for transmit diversity and spatial multiplexing with
$\nT=\nR=2$.} \label{fig3}
\end{figure}

Whereas, in the previous section, the superiority of spatial multiplexing relative
to transmit diversity was illustrated in the
context of modern wireless systems with powerful outer coding,
an opposite but incorrect conclusion can be reached if one
compares the error probabilities of the two schemes in the absence of outer coding.

Consider $\nT = \nR = 2$ for the sake of specificity. Comparisons
must be conducted at equal $\SNR$ and rate, e.g.,
Alamouti with $16$-QAM and spatial multiplexing with $4$-QAM.
These constitute two different space-time modulation formats,
both with $4$ bits per MIMO symbol.  Fig. \ref{fig3} presents the
symbol error probabilities, averaged over the fading distribution,
for a maximum-likelihood detector with no outer coding.
The difference in slopes is explained by the classical notion of diversity order:
for Alamouti the probability of error decreases as $\SNR^{-4}$, whereas
for spatial multiplexing it decreases only as $\SNR^{-2}$.
Based on these curves, one might conclude that the schemes are roughly
equivalently at low and moderate $\SNR$ and that Alamouti is markedly
superior at high $\SNR$.

How is the comparison of uncoded error probabilities to be
reconciled with the mutual-information-based comparison, where
spatial multiplexing was found to be decidedly better? The answer
lies in the outer code.

In unfaded channels, coding effectively provides a simple horizontal shift of the error probability curve.
In fading channels, however, the effect of coding is considerably more
important: not only does it provide such horizontal shift,
but it also collects diversity over the entire range of symbols spanned by each coded block \cite{Barry}.
In a system such as the one described in Tables \ref{table:parameters} and \ref{table:pdp}, the outer code makes use of frequency
selectivity across tones and time selectivity across H-ARQ rounds. Without an outer code, on the other hand,
this selectivity would not be exploited and thus, as the example in Fig \ref{fig3} attests, averaging uncoded error probabilities does not have the same
operational meaning of averaging mutual informations. Since modern communication systems rely
on powerful channel codes, inferring their performance on the basis of uncoded error probabilities can be a rather
misleading proposition.

%%%%%%%%%%%%%%%%%%%%%%%%%%%%%%%%%%%%%%%%%%%%%%%%%%%%%%%%%%%%%5
\section{Conclusion}
\label{section:conclusion}

Since the 1970's, antenna diversity had been a preferred weapon used by mobile wireless systems
against the deleterious effect of fading. While narrowband channelizations and non-adaptive links were the norm,
antenna diversity was highly effective. In modern systems, however, this is no longer the case.
Link adaptivity and scheduling have rendered transmit diversity undesirable for low-velocity users whereas
abundant time/frequency selectivity has rendered transmit diversity superfluous
for high-velocity users.
Moreover, the prevalence of MIMO has opened the door for a much more effective use of antennas: spatial multiplexing.
Indeed, the spatial degrees of freedom created by MIMO should be regarded as additional 'bandwidth' and, for the same
reason that schemes based on time/frequency repetition are not used for they waste bandwidth, transmit diversity
techniques waste 'bandwidth'. %pointless.

Of all possible DMT points, therefore, the zero-diversity one stands out in importance. Techniques, even suboptimum ones, that can
provide full multiplexing are most appealing to modern wireless systems
whereas techniques that achieve full diversity order but fall short on multiplexing gain are least appealing.
Our findings further the conclusion in \cite{Barry}, where a similar point is made solely on the basis of the multiplexing gain for frequency-flat channels.
Although this conclusion has been reached on the premise that the coded error probabilities of discrete constellations are well approximated
by the mutual information outages of Gaussian codebooks,
we expect it to hold in any situation where the code operates at a (roughly) constant gap to the mutual information.

The trend for the foreseeable future is a sustained increase in system bandwidth, which is bound to only shore up the
above conclusion. LTE, which for our case study was taken to use 10 MHz,
is already moving towards 20 MHz channelizations.

At the same time, exceptions to the foregoing conclusion do exist. These include, for example,
control channels that convey short messages. Transmit diversity is fitting for these
channels, which do benefit from a lower error probability but lack significant time/frequency selectivity.
Other exceptions may be found in applications such as sensor networks or others where the medium access control is
non-existent or does not have %the capabilities described in Section~\ref{section:modern} (chiefly
link adaptation and retransmission mechanisms.

Our study has only required evaluating well-known techniques under realistic models and at the appropriate
operating points. Indeed, a more general conclusion that can be drawn from the discussion in this paper
is that, over time,
the evolution of wireless systems has rendered some of the traditional models and wisdoms obsolete. In particular:
\begin{itemize}
\item Frequency and time selectivity should always be properly modeled. %accounted for.
\item Performance assessments are to be made at the correct operating point, particularly in terms of error probability.
\item The assumptions regarding transmit CSI must be consistent with the regime being considered. At low velocities, adaptive rate control based on instantaneous CSI
  should be incorporated; at high velocities, only adaptation to average channel conditions should be allowed.
\item Coded block error probabilities or mutual information outages, rather than uncoded error probabilities, should be used to gauge performance.
\end{itemize}

Proper modeling is essential in order to evaluate the behavior of transmission and
reception techniques in contemporary and future wireless systems. As our discussion on transmit diversity and spatial
multiplexing demonstrates, improper modeling can lead to misguided perceptions and fictitious gains.

\section{Acknowledgments}

The authors gratefully acknowledge comments, suggestions and discussions with Profs. Jeff Andrews and Robert Heath Jr., from the University of Texas at Austin,
Prof. Uri Erez, from Tel-Aviv University, Prof. Andrea Goldsmith, from Stanford University, Prof. Albert Guillen, from the University of Cambridge,
Prof. Daniel P. Palomar, from the Hong Kong University of Science and Technology, and Mr. Chris Ramming.


\begin{thebibliography}{10}

\bibitem{bohn1930}
W.~C Bohn,
\newblock ``Automatic selection of receiving channels,''
\newblock {\em U. S. Patent No. 1747218}, Feb. 1930.

\bibitem{beverage1931}
H.~H. Beverage and H.~O. Peterson,
\newblock ``Diversity receiving system of {RCA} for radiotelegraphy,''
\newblock {\em Proc. of IRE}, vol. 19, pp. 531--584, Apr. 1931.

\bibitem{altman1956}
F.~J. Altman and W.~Sichak,
\newblock ``A simplified diversity communication system for beyond-the-horizon
  links,''
\newblock {\em IRE Trans. Comm. Systems}, vol. 4, pp. 50--55, Mar. 1956.

\bibitem{bello1963}
P.~Bello and B.~Nelin,
\newblock ``Predetection diversity combining with selectively fading
  channels,''
\newblock {\em IRE Trans. Comm. Systems}, vol. 9, pp. 32--42, Mar. 1963.

\bibitem{makino1967}
H.~Makino and K.~Morita,
\newblock ``Design of space diversity receiving and transmitting systems for
  line-of-sight microwave links,''
\newblock {\em IRE Trans. Comm. Tech.}, vol. 15, pp. 603--614, Aug. 1967.

\bibitem{kahn1954}
L.~R. Kahn,
\newblock ``Ratio squarer,''
\newblock {\em Proc. IRE}, vol. 42, pp. 1704, Nov. 1954.

\bibitem{Barnett1970}
W.~T. Barnett,
\newblock ``Microwave line-of-sight propagation with and without frequency
  diversity,''
\newblock {\em Bell Syst. Tech. Journal}, vol. 49, no. 8, pp. 1827--1871, Oct.
  1970.

\bibitem{brennan1959}
D.~G. Brennan,
\newblock ``Linear diversity combining techniques,''
\newblock {\em Proc. IRE}, vol. 47, pp. 1075--1102, 1959.

\bibitem{schwartz1965}
M.~Schwartz, W.~R. Bennett, and S.~Stein,
\newblock {\em Communication Systems and Techniques},
\newblock McGraw-Hill, New York, 1965.

\bibitem{vigants1968}
A.~Vigants,
\newblock ``Space diversity performance as a function of antenna separation,''
\newblock {\em IEEE Trans. Comm. Tech.}, vol. 16, no. 6, pp. 831--836, Dec.
  1968.

\bibitem{jakes}
William~C. Jakes,
\newblock {\em Microwave Mobile Communications},
\newblock New York, IEEE Press, 1974.

\bibitem{cox83}
D.~C. Cox,
\newblock ``Antenna diversity performance in mitigating the effects of portable
  radiotelephone orientation and multipath propagation,''
\newblock {\em IEEE Trans. Commun.}, vol. 31, pp. 620--628, May 1983.

\bibitem{pdc}
N.~Nakajima K.~Kinoshita, M.~Kuramoto,
\newblock ``Development of a \mbox{TDMA} digital cellular system based on a
  \mbox{Japanese} standard,''
\newblock {\em Proc. of IEEE Vehic. Tech. Conf. (VTC'91)}, pp. 642--645, 1991.

\bibitem{Goldsmith_Wireless}
A.~J. Goldsmith,
\newblock {\em Wireless Communications},
\newblock Cambridge, 2005.

\bibitem{PSTD}
A.~Hiroike and F.~Adachi,
\newblock ``Combined effects of phase sweeping transmitter diversity and
  channel coding,''
\newblock {\em IEEE Trans. Veh. Technol.}, vol. 33, pp. 37--43, Feb. 1984.

\bibitem{wittneben}
A.~Wittneben,
\newblock ``A new bandwidth efficient transmit antenna modulation diversity
  scheme for linear digital modulation,''
\newblock {\em Proc. IEEE Int'l Conf. on Commun. (ICC'93)}, vol. 3, pp.
  1630--1634, 1993.

\bibitem{alamouti}
S.~M. Alamouti,
\newblock ``A simple transmit diversity technique for wireless
  communications,''
\newblock {\em IEEE J. Select. Areas Commun.}, vol. 16, Oct. 1998.

\bibitem{tarokh}
V.~Tarokh, N.~Seshadri, and A.~R. Calderbank,
\newblock ``Space-time codes for high data rate wireless communications:
  Performance criterion and code construction,''
\newblock {\em IEEE Trans. Inform. Theory}, vol. 44, pp. 744--765, Mar. 1998.

\bibitem{jerry}
G.~J. Foschini and M.~J. Gans,
\newblock ``On the limits of wireless communications in a fading environment
  when using multiple antennas,''
\newblock {\em Wireless Personal Communications}, pp. 315--335, 1998.

\bibitem{jerry2}
G.~J. Foschini,
\newblock ``Layered space-time architecture for wireless communications in a
  fading environment when using multi-element antennas,''
\newblock {\em Bell Labs Tech. J.}, pp. 41--59, 1996.

\bibitem{telatar}
I.~E. Telatar,
\newblock ``Capacity of multi-antenna \mbox{Gaussian} channels,''
\newblock {\em Eur. Trans. Telecom}, vol. 10, pp. 585--595, Nov. 1999.

\bibitem{cioffi}
G.~Raleigh and J.~M. Cioffi,
\newblock ``Spatio-temporal coding for wireless communications,''
\newblock {\em IEEE Trans. Commun.}, vol. 46, no. 3, pp. 357--366, Mar. 1998.

\bibitem{LTEbook}
S.~Sesia, I.~Toufik, and M.~Baker (Editors),
\newblock {\em The {UMTS} {Long Term} {Evolution}: From Theory to Practice},
\newblock Wiley, 2009.

\bibitem{AndrewsBook}
J.~G. Andrews, A.~Ghosh, and R.~Muhamed,
\newblock {\em Fundamentals of WiMAX},
\newblock Prentice Hall, 2007.

\bibitem{TseViswanathBook}
D.~Tse and P.~Viswanath,
\newblock {\em Fundamentals of wireless communication},
\newblock Cambridge University Press, 2005.

\bibitem{Gallager_LDPC}
R.~Gallager,
\newblock ``Low-density parity-check codes,''
\newblock {\em IEEE Trans. Inform. Theory}, vol. 8, no. 1, pp. 21--28, Jan.
  1962.

\bibitem{berrou}
C.~Berrou, A.~Glavieux, and P.~Thitimajshima,
\newblock ``Near \mbox{Shannon} limit error correcting coding and decoding:
  Turbo-codes,''
\newblock {\em Proc. of IEEE Int'l Conf. in Communic. (ICC'93)}, pp.
  1064--1070, May 1993.

\bibitem{Forney_Costello07}
G.~D. Forney and D.~J. Costello,
\newblock ``Channel coding: The road to channel capacity,''
\newblock {\em Proceedings of IEEE}, vol. 95, no. 6, pp. 1150--1177, June 2007.

\bibitem{andrea2}
A.~J. Goldsmith,
\newblock ``The capacity of downlink fading channels with variable rate and
  power,''
\newblock {\em IEEE Trans. Veh. Technol.}, vol. 46, pp. 569--580, Aug. 1997.

\bibitem{Costello98}
D.~Costello, J.~Hagenauer, H.~Imai, and S.~B. Wicker,
\newblock ``Applications of error-control coding,''
\newblock {\em IEEE Trans. Inform. Theory}, vol. 44, no. 6, pp. 2531--2560,
  Oct. 1998.

\bibitem{JSAClimitedfeedback}
D.~J. Love, R.~W.~Heath Jr., V.~K.~N. Lau, D.~Gesbert, B.~D. Rao, and
  M.~Andrews,
\newblock ``An overview of limited feedback in wireless communication
  systems,''
\newblock {\em IEEE J. Sel. Areas in Comm.}, To appear, 2008.

\bibitem{Hochwald_Hardening04}
B.~M. Hochwald, T.~L. Marzetta, and V.~Tarokh,
\newblock ``Multiple-antenna channel hardening and its implications for rate
  feedback and scheduling,''
\newblock {\em IEEE Transactions on Information Theory}, vol. 50, no. 9, pp.
  1893--1909, Sept. 2004.

\bibitem{Viswanathan_Tcom04}
H.~Viswanathan and S.~Venkatesan,
\newblock ``The impact of antenna diversity in packet data systems with
  scheduling,''
\newblock {\em IEEE Trans. on Communications}, vol. 52, pp. 546--549, Apr.
  2004.

\bibitem{mercuryjournal}
A.~Lozano, A.~M. Tulino, and S.~Verd\'u,
\newblock ``Power allocation over parallel {Gaussian} channels with arbitrary
  input distributions,''
\newblock {\em IEEE Trans. Inform. Theory}, vol. 52, no. 7, pp. 3033--3051,
  July 2006.

\bibitem{ozarow}
L.~Ozarow, S.~Shamai, and A.~D. Wyner,
\newblock ``Information theoretic considerations for cellular mobile radio,''
\newblock {\em IEEE Trans. Veh. Technol.}, vol. 43, pp. 359--378, May 1994.

\bibitem{CaTaBi}
G.~Caire, G.~Taricco, and E.~Biglieri,
\newblock ``Optimum power control over fading channels,''
\newblock {\em IEEE Trans. Inform. Theory}, vol. 45, pp. 1468--1489, July 1999.

\bibitem{PrasadVaranasi_IT06}
N.~Prasad and M.~K. Varanasi,
\newblock ``Outage theorems for {MIMO} block-fading channels,''
\newblock {\em IEEE Trans. Inform. Theory}, vol. 52, no. 12, pp. 5284--5296,
  Dec. 2006.

\bibitem{divmul}
L.~Zheng and D.~Tse,
\newblock ``Diversity and multiplexing: A fundamental tradeoff in multiple
  antenna channels,''
\newblock {\em IEEE Trans. Inform. Theory}, vol. 49, pp. 1073--1096, May 2003.

\bibitem{CoronelBolcskei07}
P.~Coronel and H~. Bolcskei,
\newblock ``Diversity-multiplexing tradeoff in selective-fading {MIMO}
  channels,''
\newblock {\em Proc. of IEEE Int'l Symp. on Inform. Theory (ISIT'07)}, June
  2007.

\bibitem{Narasimhan06}
R.~Narasihman,
\newblock ``Finite-{SNR} diversity-multiplexing tradeoff for correlated
  {Rayleigh} and {Rician} {MIMO} channels,''
\newblock {\em IEEE Trans. Inform. Theory}, vol. 52, no. 9, pp. 3956--3979,
  Sept. 2006.

\bibitem{AzarianElGamal07}
K.~Azarian and H.~El Gamal,
\newblock ``The throughput-reliability tradeoff in block-fading {MIMO}
  channels,''
\newblock {\em IEEE Trans. Inform. Theory}, vol. 53, no. 2, pp. 488--501, Feb.
  2007.

\bibitem{Barry}
B.~Varadarajan and J.~R. Barry,
\newblock ``The outage capacity of linear space-time codes,''
\newblock {\em IEEE Trans. Wireless Communications}, vol. 6, no. 6, pp.
  2642--2648, Nov. 2005.

\bibitem{Caire_Tuninetti}
G.~Caire and D.~Tuninetti,
\newblock ``The throughput of hybrid-{ARQ} protocols for the {Gaussian}
  collision channel,''
\newblock {\em IEEE Trans. Inform. Theory}, vol. 47, no. 5, pp. 1971--1988,
  Jul. 2001.

\bibitem{jerry3}
G.~J. Foschini, G.~D. Golden, R.~A. Valenzuela, and P.~W. Wolnianski,
\newblock ``Simplified processing for high spectral efficiency wireless
  communication employing multi-element arrays,''
\newblock {\em IEEE J. Select. Areas Commun.}, vol. 17, no. 11, pp. 1841--1852,
  Nov. 1999.

\bibitem{PARC}
S.~T. Chung, A.~Lozano, H.~C. Huang, A.~Sutivong, and J.~M. Cioffi,
\newblock ``Approaching the {MIMO} capacity with {V-BLAST}: theory and
  practice,''
\newblock {\em EURASIP Journal on Applied Signal Processing (special issue on
  MIMO)}, vol. 2002, pp. 762--771, May 2004.

\bibitem{shinlee}
H.~Shin and J.~H. Lee,
\newblock ``Capacity of multiple-antenna fading channels: Spatial fading
  correlation, double scattering and keyhole,''
\newblock {\em IEEE Trans. Inform. Theory}, vol. 49, pp. 2636--2647, Oct. 2003.

\bibitem{poweroffset}
A.~Lozano, A.~M. Tulino, and S.~Verdu,
\newblock ``High-{SNR} power offset in multiantenna communication,''
\newblock {\em IEEE Trans. Inform. Theory}, vol. 51, no. 12, pp. 4134--4151,
  Dec. 2005.

\bibitem{ImpactCorr}
A.~M. Tulino, A.~Lozano, and S.~Verdu,
\newblock ``Impact of correlation on the capacity of multiantenna channels,''
\newblock {\em IEEE Trans. Inform. Theory}, vol. 51, pp. 2491--2509, July 2005.

\bibitem{HochwaldBrink03}
B.~M. Hochwald and S.~ten Brink,
\newblock ``Achieving near-capacity on a multiple-antenna channel,''
\newblock {\em IEEE Transactions on Communications}, vol. 51, no. 3, pp.
  389--399, Mar. 2003.

\bibitem{PerfectCodes}
F.~Oggier, J.-C. Belfiore, and E.~Viterbo,
\newblock ``Cyclic division algebras: A tool for space-time coding,''
\newblock {\em Foundations and Trends in Communications and Information
  Theory}, vol. 4, no. 1.

\end{thebibliography}
\end{document}